\documentclass[twocolumn,amsmath,amssymb,aps,prl,nofootinbib,superscriptaddress]{revtex4}
\usepackage{graphicx}
\usepackage{braket}
\usepackage{amsmath}
\DeclareMathOperator{\Tr}{Tr}

\newcommand{\nuest}{2.35}
\newcommand{\nuerr}{0.21}
\newcommand{\hcest}{5.06}
\newcommand{\hcerr}{0.09}

\newcommand{\pnuest}{2.13}
\newcommand{\pnuerr}{0.15}

\begin{document}
\title{Many-body Localization Transition: Schmidt Gap, Entanglement Length \& Scaling}

\author{Johnnie Gray}
\affiliation{Department of Physics and Astronomy, University College London,
Gower Street, London WC1E 6BT, United Kingdom}
\email{john.gray.14@ucl.ac.uk}
\author{Sougato Bose}
\affiliation{Department of Physics and Astronomy, University College London,
    Gower Street, London WC1E 6BT, United Kingdom}
\author{Abolfazl Bayat}
\affiliation{Department of Physics and Astronomy, University College London,
    Gower Street, London WC1E 6BT, United Kingdom}
\date{\today}

\begin{abstract}
Many-body localization has become an important phenomenon for illuminating a potential rift between non-equilibrium quantum systems and statistical mechanics.
However, the nature of the transition between ergodic and localized phases in models displaying many-body localization is not yet well understood.
Assuming that this is a continuous transition, analytic results show that the length scale should diverge with a critical exponent $\nu \ge 2$ in one dimensional systems.
Interestingly, this is in stark contrast with all exact numerical studies which find $\nu \sim 1$.
We introduce the Schmidt gap, new in this context, which scales near the transition with a exponent $\nu > 2$ compatible with the analytical bound.
We attribute this to an insensitivity to certain finite size fluctuations, which remain significant in other quantities at the sizes accessible to exact numerical methods.
Additionally, we find that a physical manifestation of the diverging length scale is apparent in the entanglement length computed using the logarithmic negativity between disjoint blocks.
\end{abstract}

\maketitle


\begin{figure}[t]
    \centering
    \includegraphics[width=\linewidth]{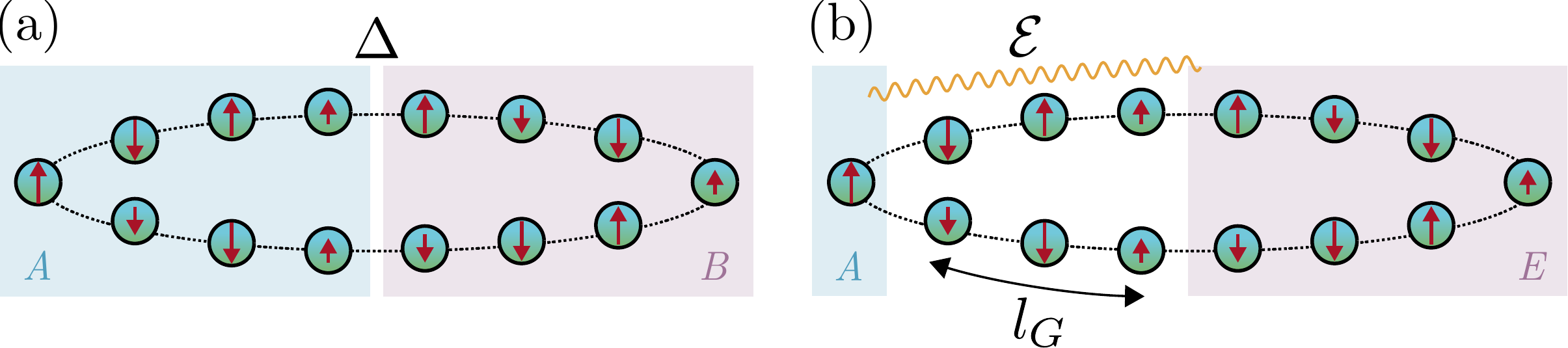}
    \caption{Schematic of the two main quantities studied here: a) the Schmidt gap, $\Delta$, across a bipartition of the system and b) the logarithmic negativity, ${\mathcal{E}}$, between disjoint blocks separated by length $l_G$.}
    \label{fig:chain}
\end{figure}

\emph{Introduction.--} It has become apparent that the Anderson Localization~\cite{anderson_absence_1958} of disordered models can survive in the presence of interactions~\cite{basko_metalinsulator_2006}, with rigorous proof now found for 1D systems~\cite{imbrie_diagonalization_2016,imbrie2016many}.
This phenomena, known as many-body localization (MBL), has attracted much interest~\cite{luitz_ergodic_2016, nandkishore_many-body_2015, huse_phenomenology_2014,sims_many-body_2013,bardarson_unbounded_2012,pal_many-body_2010}
in fundamental physics due to the fact that such systems generically break ergodicity and fail to thermalise --- thus lying beyond the scope of statistical mechanics.
Additionally, MBL occurs throughout the energy spectrum, implying that its fingerprint can be observed at all temperatures.
These facts combined have significant practical implications for quantum transport~\cite{basko_metalinsulator_2006} and information storage~\cite{yao_many-body_2015,vasseur_quantum_2015,serbyn_interferometric_2014}.
Experimental advances have allowed the controlled observation of MBL phenomena~\cite{schreiber_observation_2015,smith_many-body_2016}, further driving interest.

Considerable progress has been made in understanding the strongly localized phase, particularly in terms of local integrables of motion
~\cite{serbyn_local_2013,huse_phenomenology_2014,nandkishore_spectral_2014,chandran_constructing_2015,ros_integrals_2015,monthus_many-body_2016}
, which permit a MPS description of all eigenstates
~\cite{friesdorf_many-body_2015,serbyn_power-law_2016,wahl_entire_2016,zhang_density-matrix_2016,znidaric_diffusive_2016,devakul_obtaining_2017}.
However, eigenstates in the ergodic phase generally have volume law entanglement, restricting one to exact diagonalization techniques and small system sizes (up to $\sim20$ spins) --- this has constrained the development of a clear picture of the nature of the transition from ergodic to MBL (the MBLT).
For example, questions that still require attention include:
(i)
Which quantities can best characterize the transition?
(ii)
Is it valid to treat the MBLT using the same framework, based on the emergence of a diverging length-scale, developed for zero-temperature quantum phase transitions?
(iii)
If so, what is the universal critical exponent, $\nu$, governing this length-scale?
And (iv)
what is the physical picture of the said length-scale?

An extensive exact numerical analysis of the MBLT, using a variety of quantities, can be found in~\cite{luitz_many-body_2015}, in which finite size scaling analysis throughout the spectrum allows the observation of a mobility edge.
In fact, it is now commonplace to diagnose the MBLT with the mean energy level statistics and the block entanglement entropy~\cite{pal_many-body_2010,nandkishore_many-body_2015,luitz_many-body_2015,ponte_many-body_2015,khemani_critical_2016,khemani_two_2017,jian_solvable_2017}.
These works are largely based on the assumption that the MBLT is continuous, and their exact numerical analyses have consistently found $\nu \sim 1$.
This is in striking contrast with analytic results, found by Chayes-Chayes-Fisher-Spencer~\cite{chayes_finite-size_1986} and Chandran-Laumann-Oganesyan~\cite{chandran_finite_2015}, which would demand $\nu \ge 2/d$ for system dimension $d$ (the CCFS/CLO bound).
A recent explanation~\cite{khemani_two_2017} posits that at the finite system sizes available for exact studies, the fluctuations in these quantities are not yet dominated by the true disorder.
Thus it is highly desirable to use a new quantity better able to capture the real disorder induced transition properties.

In this letter, we bring in new tools to understand the nature of the MBLT.
Firstly, the Schmidt gap, which has been successfully employed as an order parameter in quantum phase transitions~\cite{de_chiara_entanglement_2012,lepori_scaling_2013,bayat_order_2014-1}.
Secondly, an entanglement length computed from the logarithmic negativity~\cite{peres_separability_1996,horodecki_separability_1996,lee_partial_2000,vidal_computable_2002,plenio_logarithmic_2005}, quantifying the bipartite entanglement between two disjoint blocks~\cite{wichterich_scaling_2009,wichterich_universality_2010,nobili_entanglement_2015,coser_towards_2016,coser_spin_2016-1}, which has been previously used to probe the extension of the Kondo screening cloud~\cite{bayat_negativity_2010,bayat_entanglement_2012}.
We find that, unlike previously used quantities, the Schmidt gap reveals a critical exponent $\nu\ge 2$, consistent with the CCFS/CLO bound, though, curiously as opposed to previous studies, it does \emph{not} act as an order parameter.
Moreover, we find that the  entanglement length witnesses the emergence of a diverging length scale at the transition from ergodic to MBL phase.


\begin{figure}[t]
    \centering
    \includegraphics[width=\linewidth]{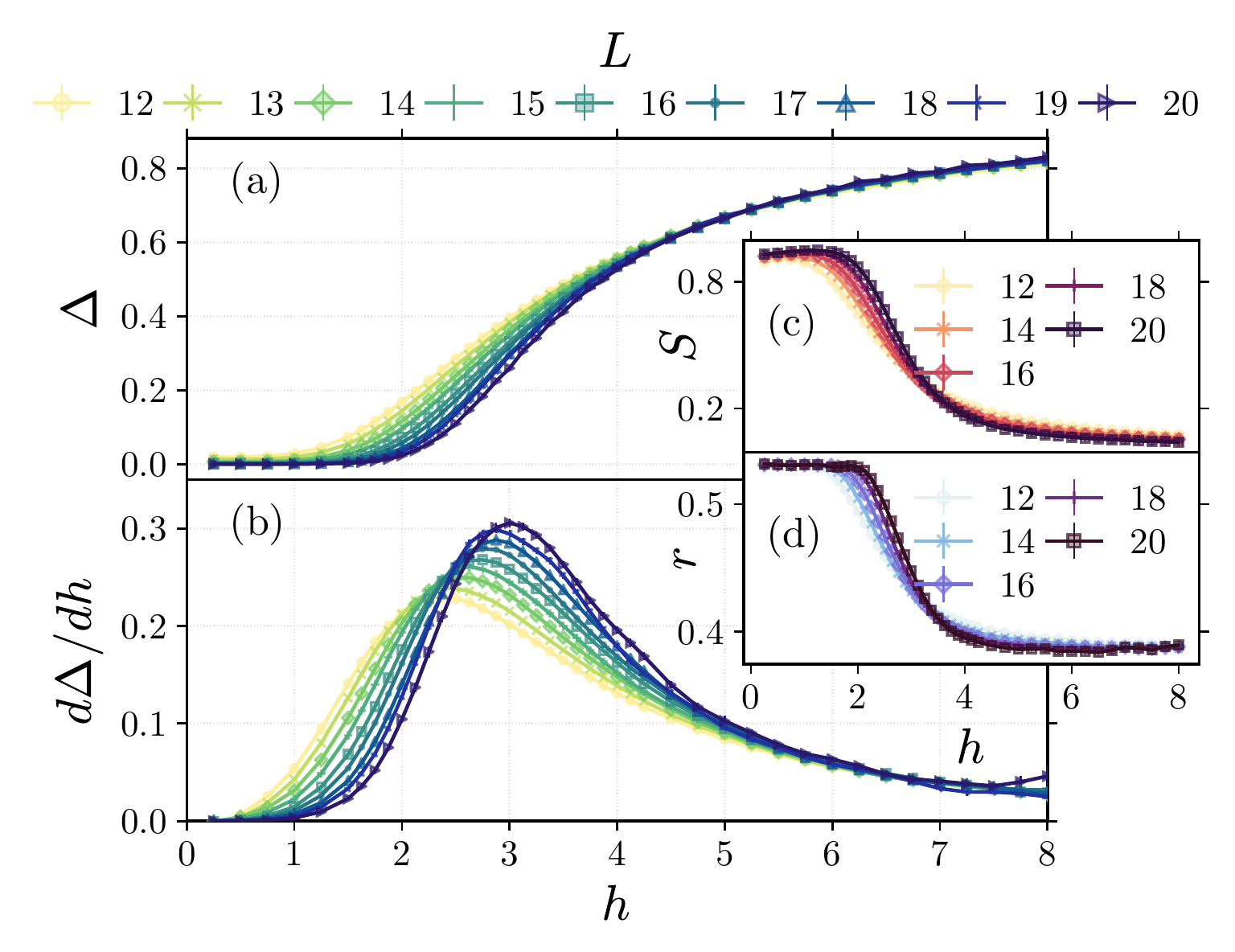}
    \caption{
    (a) and (b) :
    The Schmidt gap, $\Delta$, and its derivative as a function of disorder, $h$, across the MBLT for varying chain length, $L$.
    (c) and (d) :
    The normalized half chain entropy, $S$, and the mean energy level spacing ratio, $r$, as a function of disorder for varying $L$.
    Error bars shown where visible.
    }
    \label{fig:standard_quants}
\end{figure}

\emph{Model.--} We consider a periodic spin-$1/2$ Heisenberg chain, with random magnetic fields in the $z$-direction:
\begin{equation}
    H = \sum_{i=1}^{L} \left( J \boldsymbol{S}_i \cdot \boldsymbol{S}_{i+1}
    - h_i S^{z}_i \right),
\end{equation}
with $J$ the exchange coupling, $\boldsymbol{S}_i=\frac{1}{2} \left(\sigma^x_i, \sigma^y_i, \sigma^z_i\right)$ a vector of Pauli matrices acting on spin $i$ and dimensionless parameter $h_i$ the random magnetic field at site $i$ drawn from the flat distribution $\left[-h, h\right]$.
We diagonalize the Hamiltonian in either the spin-0 or spin-$\frac{1}{2}$ subspaces for even and odd $L$ respectively.
For each random instance we extract 50 eigenvectors, $\{\ket{E_k}\}$, in the middle of the energy spectrum~\cite{dalcin_parallel_2011,hernandez_slepc:_2005}. 
Since there is evidence of a mobility edge in MBL~\cite{luitz_many-body_2015}, at least for finite sizes, this targeting sharpens any transition observed.
The choice of 50 is a reasonable compromise on numerical efficiency whilst being statistically representative. 


\emph{Characterizing the MBLT.--}
The main quantity we compute, new in the context of MBL, is the Schmidt gap.
For two chain halves (or as close to for odd $L$), $A$ and $B$, as shown in Fig.~\ref{fig:chain}(a), an eigenvector's reduced density matrix is $\rho_{A, k} = \Tr_{B}(\ket{E_k}\bra{E_k})$ for a particular sample of the random fields.
The disorder-averaged Schmidt gap is then defined as
$\Delta = \overline{\langle\lambda_1^k - \lambda_2^k\rangle_k}$, where $\lambda_1^k$, $\lambda_2^k$ refer to the largest eigenvalues of the reduced density matrix $\rho_{A,k}$, $\langle \cdot \rangle_k$ denotes average over eigenstates and $\overline{{~}\cdot{~}}$ denotes the average over many samples.
The Schmidt gap has previously been shown to act as an order parameter for quantum phase transitions~\cite{de_chiara_entanglement_2012,bayat_order_2014-1}.
We explore the possibility of using it for characterizing the MBLT.
Unlike entanglement entropy, the Schmidt gap ignores most of the spectrum of $\rho_{A,k}$, describing only the relationship between the two dominant states across the $A-B$ cut.
This is pertinent in light of the recent finding that while the Schmidt values decay polynomially in the MBL phase~\cite{serbyn_power-law_2016}, finite size corrections are stronger for small Schmidt values.
In the ergodic phase we expect strong entanglement to produce multiple, equally likely orthogonal states, thus $\Delta\sim 0$.
In the MBL phase, however, a single dominant state should appear on either side of the cut, with $\Delta$ rising towards 1 as $h \rightarrow \infty$, implying a tensor product.
This behaviour is shown in Fig.~\ref{fig:standard_quants}(a) and becomes becomes sharper with increasing $L$.
To see this more vividly, we plot the derivative of $\Delta$  with respect to $h$ in Fig.~\ref{fig:standard_quants}(b).
The derivative has a peak at $h =\tilde{h}_c$, which not only becomes more pronounced but also shifts to the right with $L$.
We infer this to be the finite size precursor to the transition point, which suggests that in the thermodynamic limit, $L \rightarrow \infty$, the derivative of the Schmidt gap diverges at the MBLT and $\tilde{h}_c$ asymptotically approaches the transition point $h_c$.

For reference, we consider the normalised half chain entropy, widely employed to herald the MBLT~\cite{pal_many-body_2010,nandkishore_many-body_2015,luitz_many-body_2015,ponte_many-body_2015,khemani_critical_2016,khemani_two_2017}.
The von Neumann entropy of subsystem $A$ is defined as $S_{\text{vn}}=-\Tr(\rho_{A, k} \log \rho_{A, k})$.
This is normalized by the Page entropy~\cite{page_average_1993}, $S_P = (1/\log2) \sum_{i=n+1}^{mn}\frac{1}{i} - \frac{m-1}{2n}$, with $m$, $n$ the Hilbert space dimensions of subsystems $A$ and $B$, yielding the disorder-averaged $S = \overline{\langle S_{\text{vn}}\rangle_k} / S_P$.
$S_P$ is the expected entropy for a subsystem of a random pure state; since these overwhelmingly have entropy that scales as their enclosed volume, $S$ gives a measure of how far $\ket{E_k}$ has departed towards area law behaviour.
In Fig.~\ref{fig:standard_quants}(c) the behaviour of $S$ across the MBLT is shown.
In the ergodic phase its value approaches $1$ (showing the volume law), whereas in the MBL phase it falls to $0$ (representing the area law), as expected.

For reference we also compute the mean energy level spacing ratio, $r$.
For energy eigenvalues $E_k$, with gaps $\delta_k = E_{n} - E_{n-1}$, this is defined as $r = \overline{\langle \min(\delta_k, \delta_{n+1}) / \max(\delta_k, \delta_{n+1}) \rangle_k}$.
In the ergodic phase, energy level repulsion yields statistics for $r$ that match those of Gaussian Orthonormal Ensemble (GOE) random matrices~\cite{atas_distribution_2013} with $r = 0.5307(1)$.
In the MBL phase however, the eigenenergies are no longer correlated, and the energy level are simply spaced according to Poisson statistics, giving $r\approx 0.38629$.
In Fig.~\ref{fig:standard_quants}(d) the behaviour of $r$ is shown across the MBLT, clearly varying between these two statistical regimes.
For all of these quantities, we average over between 10000 for $L=10$ and 1000 for $L=20$ samples of random fields, and compute errors using statistical bootstrapping across these samples.


\begin{figure}[t]
    \centering
    \includegraphics[width=\linewidth]{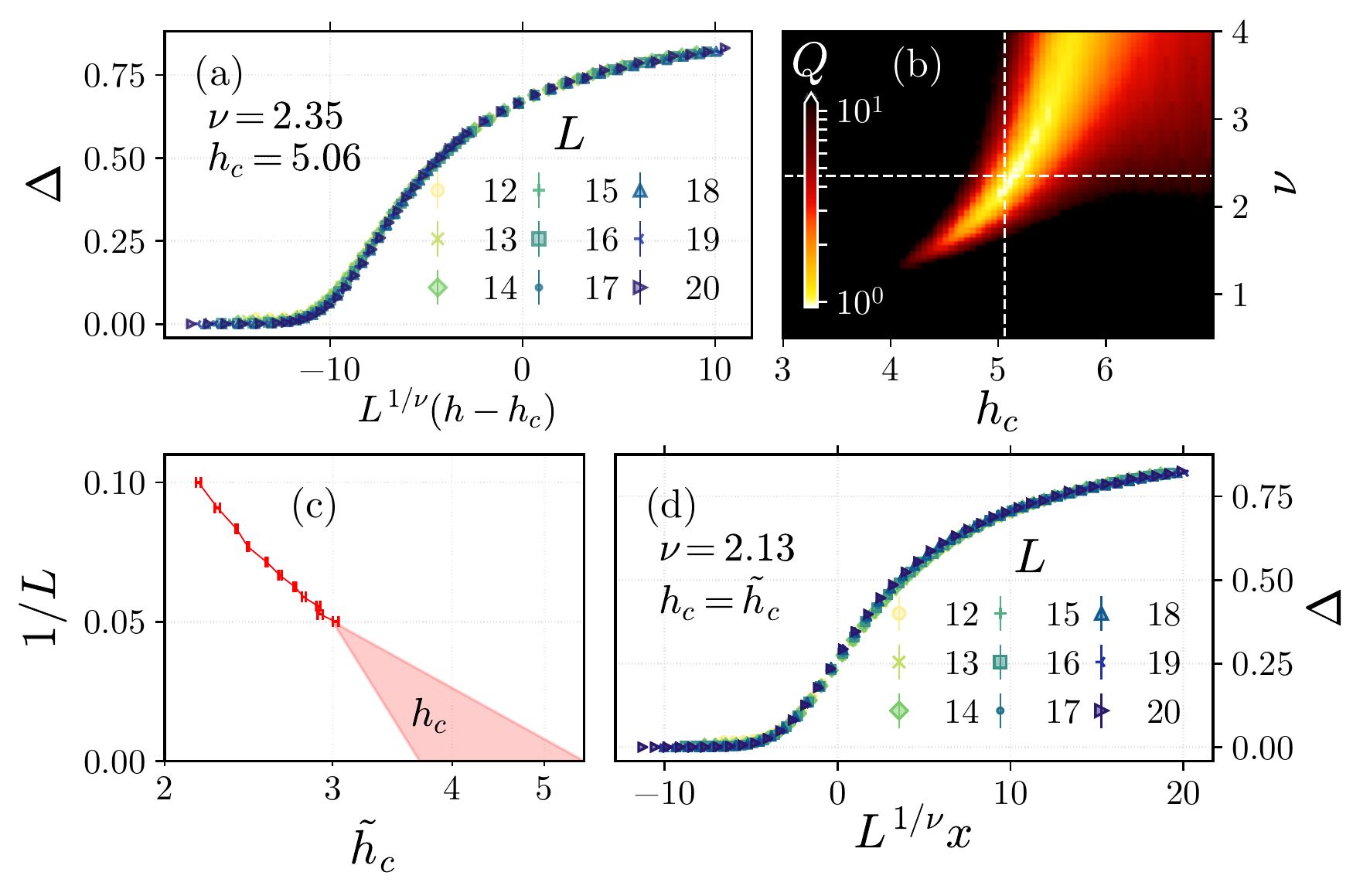}
    \caption{
    (a) Schmidt gap data collapse with fitting parameters as shown.
    (b) Quality of data collapse, $Q$, (lower is better) across the whole parameter space. The dashed lines denote the minimum point which yields the parameters shown in (a).
    (c) Pseudo-critical points $\tilde{h}_c$ as a function of inverse length $1/L$.
    (d) Schmidt gap data collapse using $\tilde{h}_c$ directly, and optimizing for $\nu$ only - the value of which is shown.
    Error bars shown where visible.
    }
    \label{fig:data_collapse}
\end{figure}

\emph{Scaling.--}
The behaviour in Figs.~\ref{fig:standard_quants}(a)-(b) suggests that the MBLT is a continuous transition in which a diverging length scale $\xi \propto |h-h_c|^{-\nu}$ emerges near the transition point, consistent with ~\cite{pal_many-body_2010}.
In order to estimate the exponent $\nu$, finite size scaling analysis~\cite{fisher_scaling_1972} has previously been employed for various quantities, including the entanglement entropy $S$.
These analyses, based on exact numerical methods, find $\nu \sim 1$~\cite{kjall_many-body_2014,luitz_many-body_2015,khemani_two_2017}, contradicting the CCFS/CLO bound.
A recently proposed explanation~\cite{khemani_two_2017}, suggests
that there are two universality classes at play here,
with that of inter-sample randomness not yet dominant for the system sizes studied.

In order to estimate $\nu$ for both models we consider the following finite size scaling ansatz,
\begin{align} \label{eq:Schmidt_Gap_ansatz}
\Delta = f(L^{1/\nu}x),
\end{align}
where $f(.)$ is an unspecified function and $x$ is ideally the scaled coordinate $h - {h}_c$.
Given the ansatz of Eq.~\ref{eq:Schmidt_Gap_ansatz}, one can then find the best fit of $h_c$ and $\nu$, using an objective function quantifying quality.
We use such a quality measure, $Q$, as refined in ~\cite{houdayer_low-temperature_2004}, which is discussed in the supplementary material.
In Fig.~\ref{fig:data_collapse}(a) we show optimal data collapse of $\Delta$
for various $L$, which is found to occur for $h_c=\hcest \pm \hcerr$ and $\nu=\nuest \pm \nuerr$.
Remarkably, this value for $\nu$ is consistent with the CCFS/CLO bound, in contrast to finite size scaling analyses for $S$ and $r$, which previous studies~\cite{kjall_many-body_2014,luitz_many-body_2015,khemani_two_2017} have generally shown to yield values of $\nu \sim 1$ -- a finding also reproduced in our analyses (data not shown).
We show the quality of collapse, $Q$, for all possible combinations of $h_c$ and $\nu$
in Fig.~\ref{fig:data_collapse}(b), the minimum point of which defines the best fit values of $\nu$ and $h_c$.
To define errors on $\nu$ and $h_c$, we perform the scaling with various subsets of data (see supplementary material) and compute the variance among all those which achieve a good quality. 

The critical $h$ we find with $\Delta$ is slightly higher than that generally reported.
One possible explanation is that a lower effective $\nu$ fits best with a lower effective $h_c$, a relation that can be seen in Fig.~\ref{fig:data_collapse}(b).
Thus it is possible that in other studies using $S$ and $r$, where $\nu \sim 1$, $h_c$ is artificially lower due to the finite size effects.
We note that a standard method of extracting $h_c$ independently -- plotting the pseudo-critical points against inverse length, shown in Fig.~\ref{fig:data_collapse}(c) -- does not give a decisive value for the real critical point.
In fact $h_c \sim 3.7$ would seem to be a lower bound on the transition point , with a value between 4.5 and 5.5 more consistent.
Additionally, if one were to identify an intersection point for all lengths in Fig.~\ref{fig:standard_quants} -- which should occur at $h=h_c$ as implied by Eq.~\eqref{eq:Schmidt_Gap_ansatz} -- this would also be at $h \sim 5$.
In contrast, the point of intersection for $S$ and $r$ shifts significantly as $L$ increases - implying a deviation from the finite size ansatz.
As a final cross-validation, to estimate $\nu$ independently from $h_c$, we take the pseudo-critical points $\tilde{h}_c$ directly to define the scaled coordinate $x$, and find the best quality of fit, $Q$, solely as a function of $\nu$.
This approach yields $\nu = \pnuest \pm \pnuerr$ -- in accordance with the first estimate -- for which data collapse is shown in Fig.~\ref{fig:data_collapse}.


\begin{figure}[t]
    \centering
    \includegraphics[width=\linewidth]{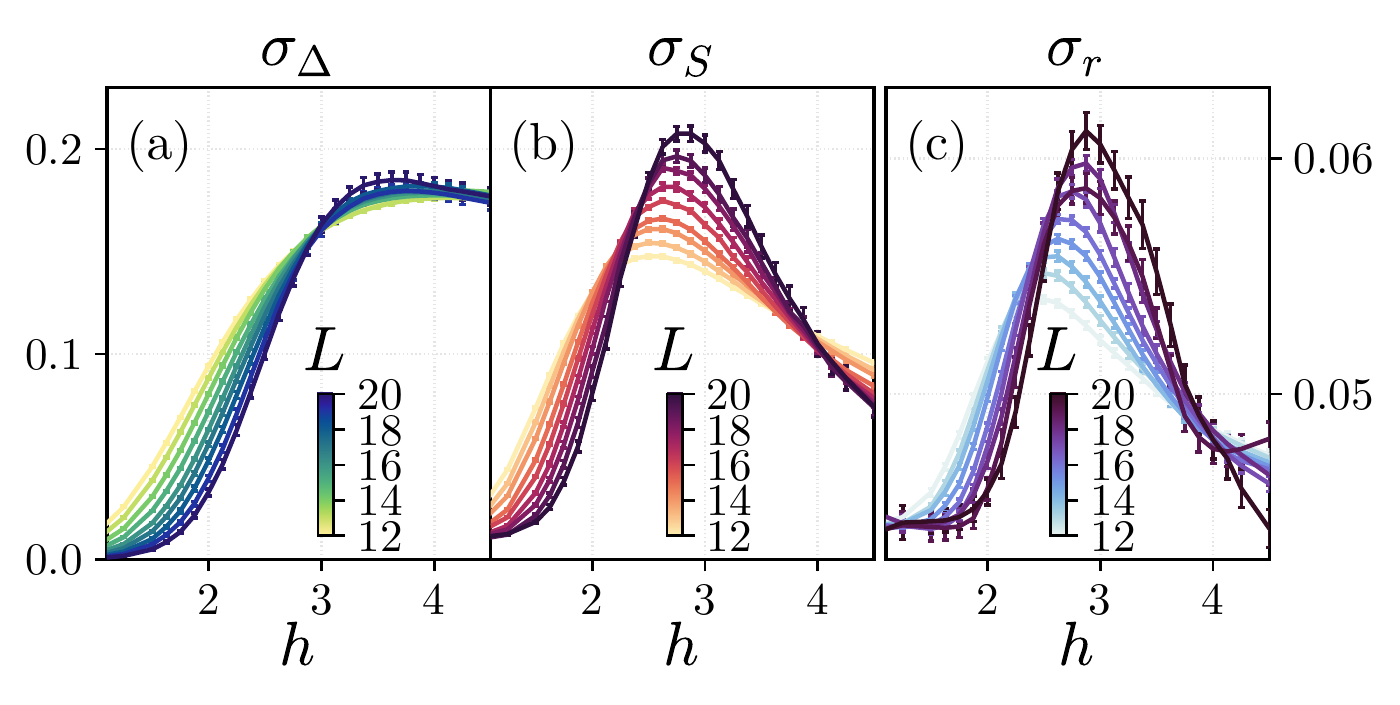}
    \caption{
        Standard deviation between samples for: (a) the Schmidt gap, $\sigma_\Delta$; (b) normalized block entropy; $\sigma_S$; and (c) mean energy level spacing ratio, $\sigma_r$. Shown as a function disorder, $h$, and length, $L$.
        Error bars shown where visible.
    }
    \label{fig:sample_variance}
\end{figure}

\emph{Sample Fluctuations.--}
In order to understand why the Schmidt gap is more successful than typical quantities, we study the fluctuation of $\Delta$, $S$ and $r$ between samples.
Motivated by Ref.~\cite{khemani_two_2017}, we consider how the size of these
fluctuations scales with $L$.
We define the standard deviations as
$\sigma_\Delta^2 = \text{Var} \left[ \langle\lambda_1^k - \lambda_2^k\rangle_k \right]$,
$\sigma_S^2 = \text{Var} \left[ \langle S_{\text{vn}} \rangle_k / S_P \right]$ and
$\sigma_r^2 = \text{Var} \left[ \langle \min(\delta_k, \delta_{n+1}) / \max(\delta_k, \delta_{n+1}) \rangle_k \right]$, with the variance
$\text{Var}[\cdot]$ taken across samples.
These are shown across the MBLT for various system sizes in Figs.~\ref{fig:sample_variance}(a-c).
All three quantities must lie between 0 and 1, thus their standard deviation is capped at 0.5.
As the figures show however, the peaks of $\sigma_S$ and $\sigma_r$ are both still rising significantly with $L$ and not yet saturated, whereas the peak of $\sigma_\Delta$ is almost constant.
The implication is that for $S$ and $r$, the effect of the small system sizes is to suppress the amount of fluctuations driven by the true disorder.
On the other hand, changing the length $L$ seems to have little effect on $\sigma_\Delta$ -- suggesting that it already experiences the full, disorder driven, thermodynamic-limit fluctuations.
A possible explanation is that finite size effects are dominantly confined to the smaller Schmidt coefficients, which still contribute significantly to $\sigma_S$.


\begin{figure}[t]
    \includegraphics[width=\linewidth]{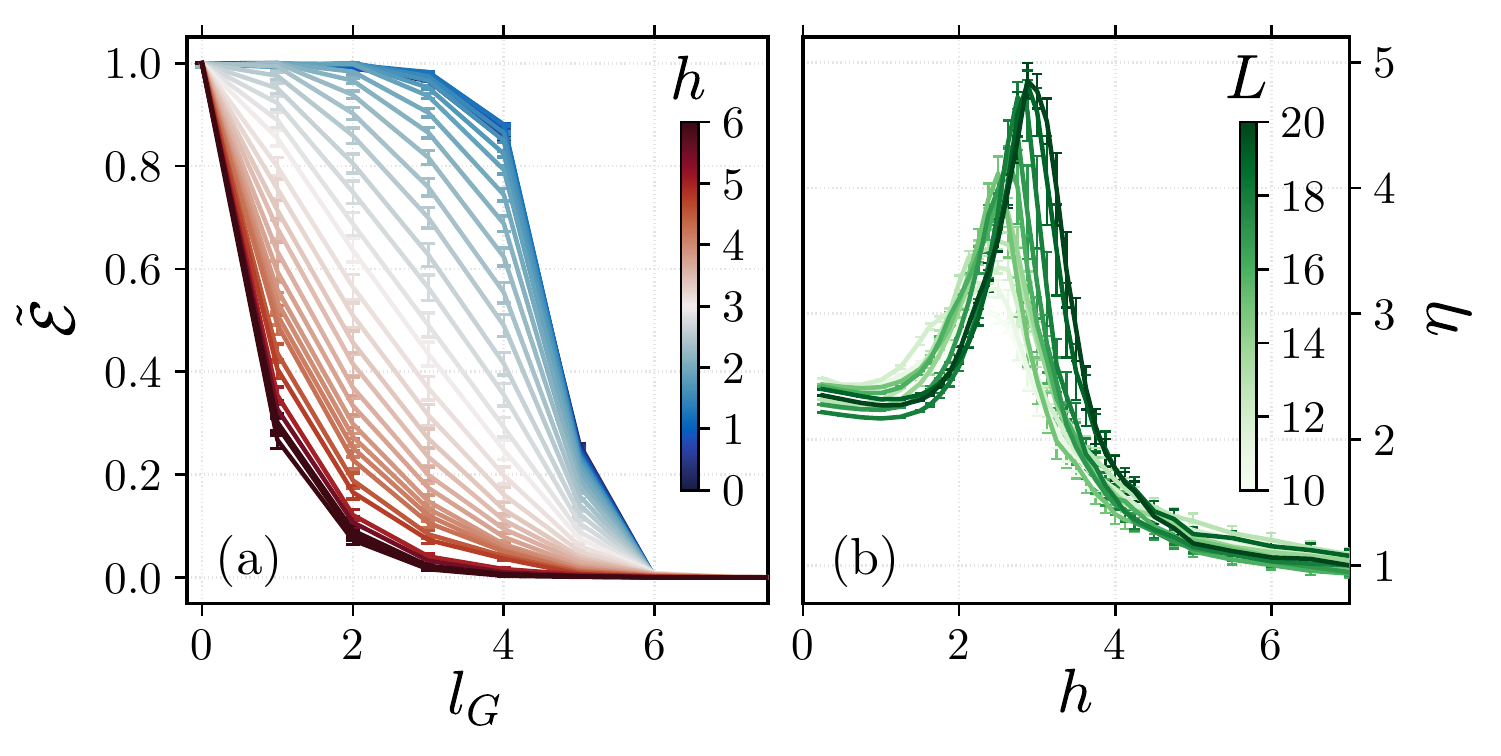}
    \caption{
        (a)
        Average logarithmic negativity as a function of the gap between two disjoint blocks as depicted in Fig.~\ref{fig:chain}, the first block being a single spin, the second the rest of the system.
        Here, $L=20$, which has a pseudo-critical point at $h \sim 3$.
        (b)
        Bipartite entanglement length, as computed with Eq.~\eqref{eq:entanglement-length}, across the MBLT for varying chain length $L$.
    }
    \label{fig:logneg_length}
\end{figure}

\emph{Entanglement length.--}  The nature of the diverging length scale $\xi$, in the context of MBLT, is mysterious and a physical picture is lacking.
To shed light on this, we introduce an entanglement length, as previously used for detecting the Kondo screening cloud~\cite{bayat_negativity_2010,bayat_entanglement_2012}.
Specifically, we consider the entanglement between a small subsystem $A$, here a single spin, and an environment $E$, separated by a gap of length $l_G$, a geometry shown in Fig.~\ref{fig:chain}(b).
The reduced state of the two blocks is $\rho_{AE,k} = \Tr_{G}(\ket{E_k}\bra{E_k})$, where $\Tr_G$ is removes the $2 l_G$ spins not in $A$ or $E$.
We use the logarithmic negativity~\cite{peres_separability_1996,horodecki_separability_1996,lee_partial_2000,vidal_computable_2002,plenio_logarithmic_2005} to quantify the entanglement between systems $A$ and $E$, defining
$\mathcal{E}(l_G)=\overline{\left\langle \log || \rho^{\Gamma}_{AE,k} ||_1 \right\rangle_k}$, with $\Gamma$ the partial transpose, and $||\cdot||_1$ the trace norm.
Since we are only concerned with the relative decay of entanglement we also define the normalized entanglement as $\tilde{\mathcal{E}}(l_G) = \mathcal{E}(l_G) / \mathcal{E}(0)$.
This naturally gives information about bipartite entanglement over a range of scales, unlike the two-site concurrence for example (which quickly goes to zero for large separation), and unlike the widely used entanglement entropy (which cannot quantify the entanglement of mixed states -- which inevitably arise when looking at two subsystems of a larger state).
In the ergodic phase, due to volume law entanglement, the eigenstates are highly multipartite entangled between their spins.
This implies that any reduced state of two small blocks is close to the identity and thus very weakly entangled.
From this two features can be inferred: i) $\tilde{\mathcal{E}}(l_G)$ is initially expected to decay slowly with increasing $l_G$, and ii) $\tilde{\mathcal{E}}(l_G)$ must go to zero as $l_G \rightarrow L/2$.
Since this precludes a linear type decay, it is expected that there is a distance at which $\tilde{\mathcal{E}}(l_G)$ rapidly decays - indeed we find this to be the case, with a sharp drop-off when half the system is traced out, i.e. $l_G \sim L / 4$.
In the MBL phase, however, $A$ will be weakly entangled with only spins close to it, and thus $\tilde{\mathcal{E}}(l_G)$ should decay quickly even for small $l_G$.
In Fig.~\ref{fig:logneg_length}(a) we plot $\tilde{\mathcal{E}}$ as a function of $l_G$ for various disorder strengths $h$ in a chain of length $L=20$.
As is clear from the figure the location of the main drop in $\tilde{\mathcal{E}}$ varies significantly with $h$.
While in the ergodic phase $\tilde{\mathcal{E}}$ this decay is concentrated at $l_G \sim L / 4$, in the MBL phase it is concentrated at $l_G \sim 1$.
Interestingly, at the pseudo-critical point, ($\tilde{h}_c \sim 3$ for $L=20$, see Fig.~\ref{fig:standard_quants}(b)), entanglement decays close to linearly --- each spin lost contributes equally to the entanglement, implying that the bipartite entanglement is equally spread over many sites.
This fits with a picture of a self-similar structure of entangled clusters~\cite{potter_universal_2015,khemani_critical_2016}.
The detailed behaviour of $\tilde{\mathcal{E}}$ as a function of system size can be found in the supplementary material.

To extract a length scale from $\tilde{\mathcal{E}}(l_G)$ we define a length, $\eta$, from the maximum inverse gradient as such:
\begin{equation}\label{eq:entanglement-length}
    \eta = \max_{l_G}|d \tilde{\mathcal{E}} / d{l_G}|^{-1}.
\end{equation}
Assuming the fastest decay is exponential-like, this quantity naturally arises from expressions of the form $\tilde{\mathcal{E}} \propto e^{-l_G/\eta}$.
This is a more robust way of finding an exponential fit in the region of the most rapid decay of $\tilde{\mathcal{E}}$, or a more general fit for the full behaviour.
At the transition point, where $\tilde{\mathcal{E}}$ decays linearly, $\eta$ takes its maximum value, since the gradient is always small, or equivalently, a very slow exponential fit is needed.

The behaviour of $\eta$ as a function of $h$ for varying $L$ is shown in Fig.~\ref{fig:logneg_length}(b), in which it can be seen to sharply peak at $h\sim\tilde{h}_c$ for each $L$ across the critical region,
-- evidence that the diverging length scale $\xi$ is closely captured by the length $\eta$.
In the supplementary material we show that taking the initial block as 2 spins yields almost identical results.
A plausible explanation for the \emph{increase} in $\eta$ as one approaches the MBLT from the ergodic side is that proximal spins become off-resonant so that bonding (\emph{bipartite} entanglement) takes place at increasingly longer scales -- a process that is not possible if the spins are part of a large multi-partite entangled block.
We note several interesting approaches that made use of the two site concurrence~\cite{bera2016local,iemini2016signatures} or mutual information~\cite{de_tomasi_quantum_2017}, which despite revealing other interesting features, such as scaling, do not show a divergence in the localization length from both sides of the transition.
An alternative approach to identifying the diverging length scale on the ergodic side based on the entanglement spectrum has been recently developed in Ref.~\cite{pietracaprina_entanglement_2016}.
It is an interesting open question whether that length is related to the entanglement length proposed here.


\emph{Conclusions.--}
In this letter we have explored the MBLT using the Schmidt gap and the entanglement length.
We show that the Schmidt gap not only exhibits scaling at the MBLT, but does so with a critical exponent $\nu>2$, compatible with analytic predictions.
This compatibility is absent in all quantities studied with exact numerical methods thus far, a fact that we attribute to the presence of significant finite size effects which the Schmidt gap is less sensitive to.
We have also considered an entanglement length computed using the logarithmic negativity across two disjoint blocks, which yields a diverging length scale at the MBLT.

\begin{acknowledgments}
\emph{Acknowledgements.-- }
JG acknowledges funding from the EPSRC Center for Doctoral Training in Delivering Quantum Technologies at UCL. AB and SB acknowledge the EPSRC grant EP/K004077/1. SB acknowledges financial support by the ERC under Starting Grant 308253 PACOMANEDIA.
\end{acknowledgments}

\clearpage
\widetext

\setcounter{equation}{0}
\setcounter{figure}{0}

\makeatletter
\renewcommand{\theequation}{S\arabic{equation}}
\renewcommand{\thefigure}{S\arabic{figure}}

\section{Supplementary Material}

\begin{figure}[!b]
    \centering
    \includegraphics[width=0.55\linewidth]{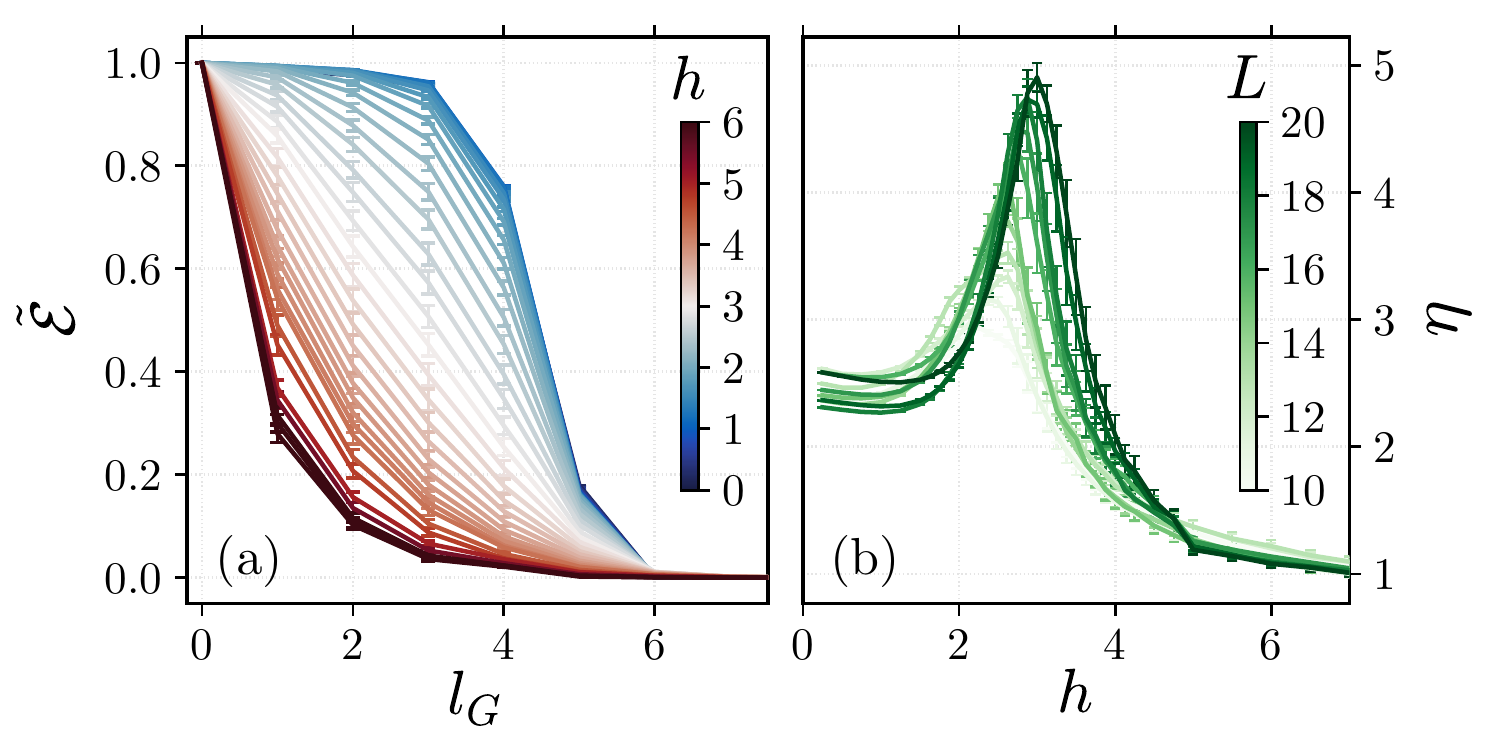}
    \caption{
            (a)
            Average logarithmic negativity as a function of the gap between two disjoint blocks, the first block being two spins, the second the rest of the system, for $L=20$, which has a pseudo-critical point at $h \sim 3$.
            (b)
            Bipartite entanglement length, as computed with Eq.~(3), across the MBLT for varying chain length $L$.
            Error bars shown where visible.
        }
    \label{fig:logneg2}
\end{figure}

\subsection{Quality of Collapse}


To find the quality of data collapse of scaled Schmidt data for a given $h_c$ and $\nu$ we use the `pyfssa' program~\cite{andreas_sorge_2015_35293}.
The underlying procedure, which is based on the method as refined in~\cite{houdayer_low-temperature_2004} is as follows.
Assume we have data points $y_{ij}$ (e.g the Schmidt gap, $\Delta$) and their standard errors $dy_{ij}$, where $i$ indexes the lengths $L_i$ and $j$ the disorder strengths $h_j$. Since we assume that there is only a correlation length exponent, $\nu$, we scale only the disorder strength as such:
$x_{ij} = L^{1/\nu}_i(h_j - h_c)$.
A curve is then fitted through the data using least squares which yields the fitted points $Y_{ij}$ and their estimated errors $dY_{ij}$.
The quality can then be defined as a $\chi^2$ statistic based on the relative deviation from this fitted curve:
\begin{equation}\label{eq:quality-collapse}
    Q = \frac{1}{\mathcal{N}} \sum_{i,j} \frac{(y_{ij} - Y_{ij})^2}{dy^2_{ij} + dY^2_{ij}}
\end{equation}
with normalization $\mathcal{N}$ accounting for the number of terms where the fitted curve is defined.
This quantity is minimized when all the actual data lies close to the curve of best fit.
In particular, when the deviations from the fit are approximately equal to the uncertainty in the fit, $Q \sim 1$, which is generally what we find with the Schmidt gap $\Delta$.

Since finite size scaling is only strictly relevant close to the transition, and finite size effects may be too strong at very small lengths, there is also some freedom in how one selects the data to scale.
Similarly to~\cite{luitz_many-body_2015}, we vary the size of the window around $h_c$ to select, the minimum $L$ of which to include data for, and also whether to include odd as well as even $L$.
The advantage of using Eq.~\eqref{eq:quality-collapse} is that the the combinations of the above which give statistically reasonable collapse can be objectively identified.
The minimum $L$ is chosen such that increasing it does not significantly change the values of $\nu$ and $h_c$ found.
One can then average over all parametrizations that achieve a `good' value of $Q$ (determined visually to be $\sim 10$) -- yielding the plot in Fig.~3(b).

The different parametrizations above, as well as bootstrap sampling over disorder realizations, then yields a spread in the locations of $h_c$ and $\nu$ where the best value of $Q$ is found. This allows an estimation of the error in $h_c$ and $\nu$ which takes into account both the collapse method and the random error.

\subsection{Entanglement Length with Larger Block Size}

For completeness, we show here the equivalent of Fig.~5 taking instead the size of the block A to be 2. As can be seen in Fig.~\ref{fig:logneg2}, this yields almost exactly the same shapes and plots, including a divergence of the length $\nu$ at the pseudo-critical points $\tilde{h_c}$, which shift right with length $L$.

\subsection{Detailed Behaviour of the Disjoint Entanglement vs. $L$}

\begin{figure*}[hb]
    \centering
    \includegraphics[width=0.9\linewidth]{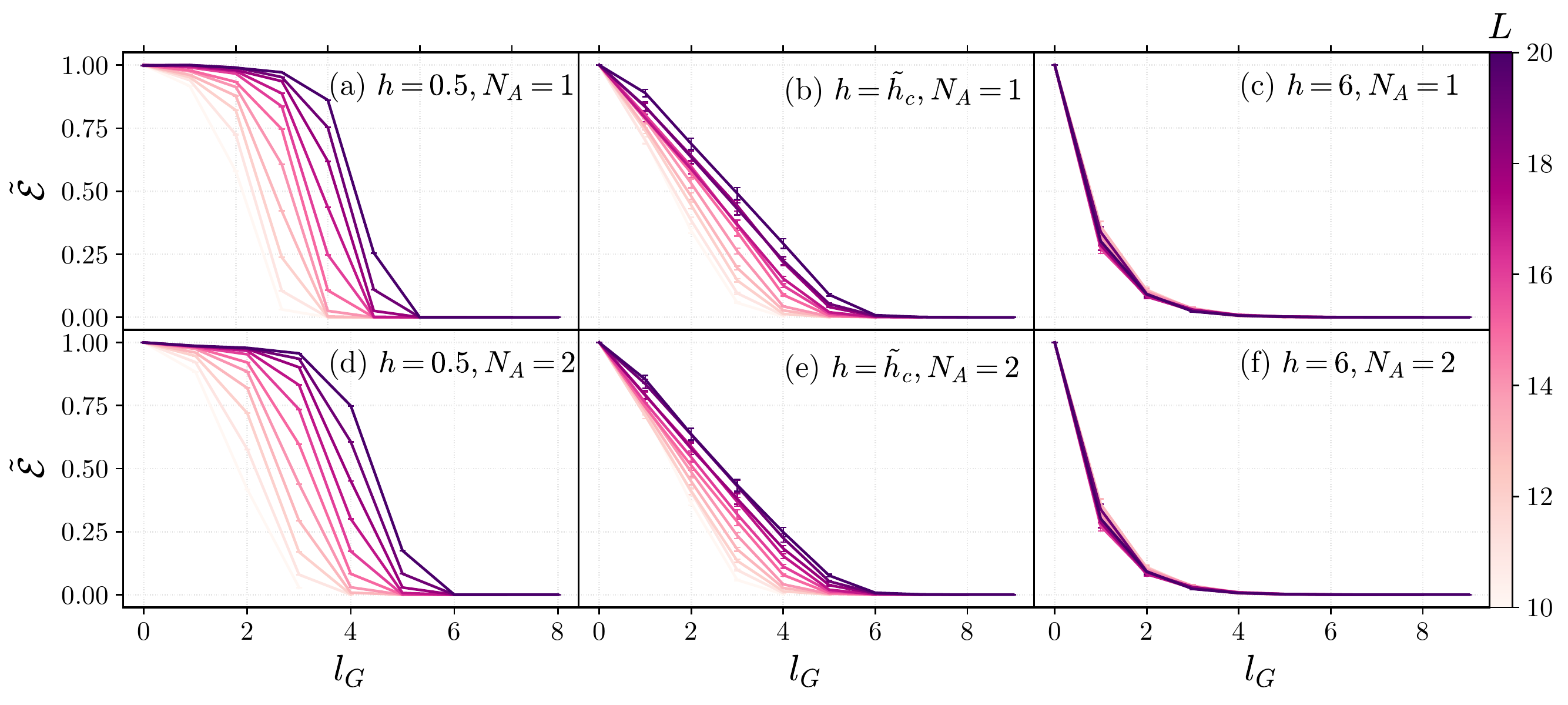}
    \caption{
    Behaviour of the normalized disjoin block entanglement, $\tilde{\mathcal{E}}$, as a function of the gap size $l_G$ for varying chain lengths $L$. The top line of panels corresponds to an initial blocksize of $N_A=1$, and the bottom 2.
    The left column is shows behaviour deep in the ergodic phase, the middle behaviour at the pseudo-critical points, and the right behaviour deep in the localized phase.
    Error bars shown where visible.
    }
    \label{fig:logneg-detailed}
\end{figure*}

Finally, in Fig.~\ref{fig:logneg-detailed}, we show in detail the behaviour of the normalized disjoint entanglement as a function of $L$, which sheds some light on the physical picture of the entanglement length.
In the ergodic phase -- left column, $h=0.5$ -- one can see that the point of decay for $\tilde{\mathcal{E}}$ shifts linearly to the right with system size $L$, but the length scale of the decay (which is what $\eta$ captures) remains constant.
In fact, the curves collapse onto each other with a shift of $-L/4$ (not shown).
In the localized phase -- right column, $h=6$ -- the decay of entanglement, and thus $\eta$, is practically identical for all the lengths.
At the transition point however -- central column, $h$ taken as pseudo-critical points $\tilde{h}_c$ -- the scale of decay stretches with system size.
Lastly, we note that the difference between the two block sizes tested (upper row 1, lower row 2) is very minimal.

\end{document}